\documentclass[review]{elsarticle}

\usepackage{hyperref}

\usepackage{amsmath,epsfig}
\usepackage{balance}

\usepackage[ruled,vlined]{algorithm2e}
\usepackage[]{graphicx}
\graphicspath{{Figures/}}
\usepackage{tabu}
\usepackage{lipsum} 
\usepackage{color}

\usepackage{lipsum}
\usepackage{url}
\usepackage{caption}
\usepackage{subfig} 
\usepackage{color}
\usepackage{balance}
\usepackage{float}
\usepackage{multirow}
\usepackage{float}
\usepackage{breqn}
\usepackage{tabularx}
\usepackage{todonotes}
\usepackage{xcolor}
\usepackage{verbatim}

\journal{Healthcare Analytics}

\begin{document}

\begin{frontmatter}






\title{Using U-Net Network for Efficient Brain Tumor Segmentation in MRI Images}

\author[add1]{Jason~Walsh}
\ead{jason.walsh3@ucdconnect.ie}
\author[add2]{Alice~Othmani}
\ead{alice.othmani@u-pec.fr}
\author[add1,add3]{Mayank~Jain}
\ead{mayank.jain1@ucdconnect.ie}
\author[add1,add3]{Soumyabrata~Dev\corref{mycorrespondingauthor} }
\cortext[mycorrespondingauthor]{Corresponding author. Tel.: + 353 1896 1797.}
\ead{soumyabrata.dev@ucd.ie}

\address[add1]{School of Computer Science, University College Dublin, Ireland}
\address[add2]{Universit\'e Paris-Est Cr\'eteil Val de Marne - Universit\'e Paris 12, France}
\address[add3]{ADAPT SFI Research Centre, Dublin, Ireland}

\begin{abstract}
Magnetic Resonance Imaging (MRI) is the most commonly used non-intrusive technique for medical image acquisition. Brain tumor segmentation is the process of algorithmically identifying tumors in brain MRI scans. While many approaches have been proposed in the literature for brain tumor segmentation, this paper proposes a lightweight implementation of U-Net. Apart from providing real-time segmentation of MRI scans, the proposed architecture does not need large amount of data to train the proposed lightweight U-Net. Moreover, no additional data augmentation step is required. The lightweight U-Net shows very promising results on BITE dataset and it achieves a mean intersection-over-union (IoU) of $89\%$ while outperforming the standard benchmark algorithms. Additionally, this work demonstrates an effective use of the three perspective planes, instead of the original three-dimensional volumetric images, for simplified brain tumor segmentation.

\end{abstract}


\end{frontmatter}

\section{Introduction}


Tumors are groups of cells which form abnormal tissue or growths within the human anatomy. Tumors can either be malignant where the growth is cancerous and will invade surrounding cells, or benign where the suspected growth is not cancerous~\cite{patel2020benign}. Manual identification of these abnormal growths in human anatomy is not only onerous but might also be be difficult from the perspective of a medical physician. Hence, the need for intelligent systems which can automatically detect the presence of cancer in a desired region of the human body~\cite{icsin2016review,md2022automatic}. With the advancement of technologies in the medical field, there has been a tremendous impact on diagnostics and predictive analysis of diseases. We observe an advancement of healthcare analysis in brain tumor segmentation, heart disease prediction~\cite{pathan2022analyzing}, stroke prediction~\cite{nwosu2019predicting,dev2022predictive}, identifying stroke indicators~\cite{pathan2020identifying}, real-time electrocardiogram (ECG) anomaly detection~\cite{sivapalan2022annet}, and amongst others.

\begin{figure}[!ht]
    \centering
      \includegraphics[scale=0.45]{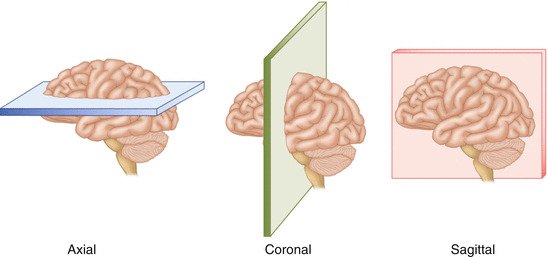}
      \caption{The three perspective planes used in medical imaging are Axial (or Transversal), Coronal and Saggital~\cite{medical_imaging}.}
    \label{fig:planes}
\end{figure}

Brain tumors are such abnormal growths found within the human cranium. Given the complex and sensitive nature of the brain, a non-invasive technology, i.e. Magnetic Resonance Imaging (MRI), is the most popular pick for brain tumor diagnosis. These images are three-dimensional scans of a patients brain and can be visualized on either of its three respecting image planes (Coronal, Sagittal and Transversal), as shown in Figure.~\ref{fig:planes}~\cite{medical_imaging}. Each perspective plane displays its information regarding a potential abnormal growth within the cranium. This classification of MRI scans based on the perspective planes has been noted to improve the analytical results while detecting brain tumors~\cite{hu2022mutual}.

Brain tumor segmentation aims to autonomously and accurately identify the size and location of a brain tumor from MRI scans. While traditional machine learning techniques require hand crafted features to perform well, most of the current research is focused on using deep learning networks to segment a region of interest (ROI) from an input image. Although considerable success has been achieved using deep learning, they either require large amounts of annotated data~\cite{liu2021review} or they depend on aggressive data augmentation techniques~\cite{ronneberger2015u}. However, a lightweight approach is almost always preferred for practical implementations~\cite{li2018lightweight,ali2022implementation}. To this end, this paper\footnote{To facilitate the reproducibility of this research, the code of this paper is made available at: \url{https://github.com/Walshj73/UNET-Tumour-Segmentation}} makes the following key contributions:


\begin{itemize}
    \item The paper demonstrates an effective use of the three perspective planes, instead of the original three-dimensional volumetric images, for simplified brain tumor segmentation
    \item A lightweight implementation of U-Net is proposed to provide accurate real-time segmentation
    \item The proposed model is systematically benchmarked with several widely used segmentation algorithms.
\end{itemize}


The remainder of this article is organized as follows. Section~\ref{sec:segTech} discusses existing image segmentation techniques that are frequently used in the domain of medical imaging. Also, in this Section we introduce the four algorithms which were chosen for benchmarking purposes in this paper. Section~\ref{sec:data} follows up with the introduction of the dataset which is used in this study, and the pre-processing operations which were performed on the images. In Section~\ref{sec:methods}, the proposed methodology is discussed in detail. The results are then discussed in Section~\ref{sec:results}, alongside a detailed comparison with the benchmarking algorithms. Lastly, Section~\ref{sec:conc} concludes the paper by summarizing the novelty, impact and obtained results.



\section{Brain Tumor Segmentation Techniques}\label{sec:segTech}

Image segmentation is a crucial area of research in the broad domains of image processing and computer vision, with applications in varied fields~\cite{dev2019cloudsegnet,dev2016color}. 
The challenge is to classify each pixel as a part of different objects in the image. Over the years, many algorithms have been proposed for this task~\cite{jain2021using,minaee2021image}. The process of segmentation has seen widespread use in the field of medical imaging as well~\cite{liu2021review}. Particularly for MRI scans, most of the previous studies modifies the existing techniques of image segmentation to manipulate the three-dimensional volumetric images~\cite{bien2018deep,zhou2018unet++}. These modified networks are further fine tuned to improve the performance over the given task.

Based on the complexity of the algorithm they use to extract the ROI from input images, segmentation techniques can be classified into $4$ broad categories. In the subsequent sub-sections, we will explain these categories and some of the corresponding algorithms that fall in each category. 

\subsection{Thresholding}

Thresholding is a very simplistic segmentation technique that is used to convert a gray-scaled or a red-green-blue (RGB) image into a binary image~\cite{pham2000current}. Although this technique is simplistic in nature, it is very effective in extracting the ROI from an image. Prior research on this particular segmentation technique explores the use of Otsu's method~\cite{Otsu1979} and global thresholding for medical image segmentation. Otsu's method is a popular segmentation algorithm used in the area of pattern recognition where features of an ROI are extracted from an image for further processing. 

However, more recent success in this domain was obtained using simple binary thresholding coupled with the watershed algorithm to segment brain tumors from MRI brain scans~\cite{mustaqeem2012efficient}. The proposed method first uses median filtering to remove various noise levels that may be damaging the quality of each image. Binary thresholding is then applied to the image, followed by the watershed algorithm to extract the ROI from the brain scan. Finally, small batches of morphological operations are used to refine the segmentation which results in a more accurate segmentation. 

\subsection{Classifiers}


Classifiers are common techniques used in segmentation which share a similar approach with methods used in supervised learning algorithms that are used in several applications~\cite{ho2021multivariate,dev2019identifying}. These algorithms learn features, which are critical to make the classification decision, from an existing dataset of annotated (or labelled) images. These labels are manually tagged to create the ground truth for learning. A simple classification algorithm which is commonly used for image segmentation is the k-nearest neighbour classifier. In this machine learning algorithm, features of each pixel are computed at first. Then the pixels with high degree of similarity within these features, are merged together~\cite{pham2000current}.

Hidden Markov Models (HMM) have been proven to perform significantly better than Support Vector Regression (SVR) models when used for brain tumor segmentation. Particular research has focused this topic on how HMM can produce a better Peak Signal-to-Noise Ratio (PSNR) and Mean Squared Error (MSE)~\cite{sharma2019improved}. The research applies a HMM to a two-dimensional MRI scan, extracted from the BITE dataset~\cite{mercier2012online}. The designed model will segment the cancerous portion of the image and produce a set of probabilities which will belong to either of the categories, \textit{i.e.} cancerous/non-cancerous. These images were further classified using sorting factor-kappa, and their segmentation performance was evaluated based on the chosen evaluation criteria.

\subsection{Clustering}

The operation of clustering is very similar to the operation of classification algorithms. Classifiers use a target dataset to perform image segmentation, whereas clustering is an unsupervised learning algorithm~\cite{jain2022internal}. In other words, there is no target labelled dataset in this case to learn the features from. Hence, unsupervised algorithms essentially train themselves by identifying the possible underlying patterns~\cite{jain2020clustering}. Commonly used clustering algorithms include k-means and the fuzzy c-means algorithm~\cite{jain2021validation}. K-means, in the domain of image segmentation, is used to segment an area of interest from the background of an input image. This method will partition the features of the dataset (pixels/voxels) into several clusters. Here, each feature will belong to a cluster with the closest mean. Fuzzy c-means or soft k-means is a variation of k-means where each feature can belong to more than one cluster based on the degree of membership~\cite{dev2014systematic}.

Prior research has explored the use of both algorithms in their ability to segment tumors from an MR image. In a particular research, Kabade and Gaikwad used advanced k-means and fuzzy c-means algorithms for segmentation~\cite{kabade2013segmentation}. Along with this, they used thresholding for feature extraction and edge detection for approximate reasoning to recognize the characteristics of the tumor. A crucial extension of this procedure was the addition of `skull stripping' where the outer cranium is extracted using the watershed algorithm to improve the segmentation accuracy of the model~\cite{abdel2015brain}. The results obtained using such methods were adequate for the dataset collected, However, they did not use three-dimensional imagery. Instead, the MRI slices were converted to a two-dimensional format and the algorithms were then applied to this new set of images.

\subsection{Deep Learning}

A Convolutional Neural Network (CNN) is a class of deep neural network which is prominently used for the analysis of two-dimensional imagery~\cite{wang2022amdcnet}. Deep learning has seen great success in medical image analysis where researchers have focused primarily on using deep learning to create systems which can accurately detect if an image contains an apparent health-related issue. MRNet is one such system~\cite{bien2018deep}. It is a convolutional neural network used to detect knee-related abnormalities.

MRNet had excellent accuracy in classifying the three knee abnormalities which were presented as inputs to the network. This network is an example of how deep learning can be used to produce systems which can accurately identify health issues from medical images. MRNet is an image classification system which is primarily used to classify if a particular MRI image contains one of the three abnormalities. The network computes this via an output probability and uses logistic regression to determine which class the probability may belong to.

MRNet was a particular architecture designed for image classification. Applying similar principles to a convolutional neural network, other network architectures, like U-Net and V-Net, were designed which could segment an ROI from an input two or three-dimensional image. In a convolutional neural network, there are multiple convolutional layers which gradually extract the features of the image through a variety of kernels and pooling layers. These networks are more actively developed when researchers explored the opportunity of using deep learning architectures for advanced image segmentation.

LinkNet proposes a deep neural network architecture similar to U-Net. The network allows for learning without any significant increase in the number of trainable/non-trainable parameters~\cite{chaurasia2017linknet}. LinkNet's success was based on its lightning-fast speed which was due to its lightweight architecture. The architecture which LinkNet utilizes holds at least $11.5$ million parameters, it is similar to U-Net, where several encoder and decoder blocks slowly break down an image and rebuild the outcome through a series of final convolutional layers. LinkNet's structure was purposely designed to minimize the total number of parameters the network contains. This allowed segmentation to be performed in real-time. This is why the network has produced a state-of-the-art performance on the Cambridge-driving Labeled Video Database (CamVid)~\cite{BrostowFC:PRL2008}.
\medskip


\section{Magnetic Resonance Image (MRI) data}\label{sec:data}

MR images can come in several different file formats, each of which shares their characteristics regarding the type of data stored within the three-dimensional image file. Digital Imaging and Communications in Medicine (DICOM) and Neuroimaging Informatics Technology Initiative (Nifti) files are some of the most common medical imaging formats widely available for research purposes. These medical images are not two-dimensional, they do not share the same characteristics as a typical two-dimensional image format such as Portable Network Graphics (PNG) or Joint Photographic Experts Group (JPEG). Medical images are three-dimensional and can be thought of as a single file which contains multiple slices of the brain over three perspective planes, where each individual slice is a collection of voxel's where each of which defines a point in three-dimensional space.

For this analysis, datasets which only contain `ground truth' were chosen for this particular image segmentation task. The cancer imaging archive~\cite{clark2013cancer} is a massive repository containing thousands of images all related to cases where the cancer was found from a pre-operative MRI scan. These datasets are normally accompanied by a post-operative scan of the affected region to show that the abnormal growth was successfully removed during surgery. Without manually extracted ground truth images it is a difficult process to perform image segmentation on a dataset containing brain tumors. There are alternatives to this issue which can be used to generate artificial pathological ground truth from a simulated dataset such as BrainWeb\footnote{McGill University, 1997. BrainWeb. www.bic.mni.mcgill.ca/brainweb/}.

\begin{figure}
    \centering
      \includegraphics[scale=0.7]{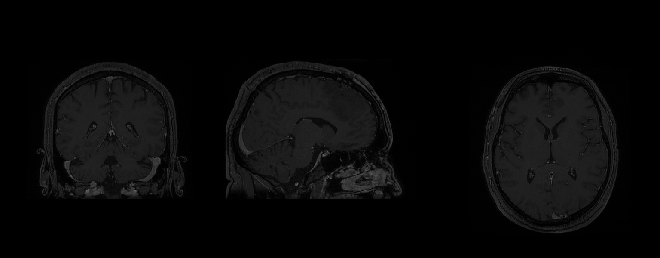}
      \caption{Anatomical plot of patient 2's pre-operative MRI scan showing all three perspective planes.}
    \label{fig:anatomical}
\end{figure}

One of the most common datasets widely available for brain tumor segmentation is the BRATS dataset for multimodal brain tumor segmentation. This dataset is widely used for biomedical image analysis and significant amounts of research have been performed using this dataset. As extensive research has been performed using this dataset we decided to look elsewhere and use the BITE's dataset~\cite{mercier2012online}. This dataset contains pre-operative, post-operative and ground truth images for 14 patients acquired by the Montreal Neurological Institute in 2010.

Each pre-operative and post-operative scan is a T1-weighted MRI accompanied by B-mode (Ultrasound) images. The images of the BITE database are split into four groups based on the specific analysis performed by the neurosurgeons. Our analysis is centred on group number three which includes 14 MR images taken before and after the patient's surgery. The dataset is stored in Medical Imaging NetCDF (MINC) format the standard format at the Montreal Neurological Institute for image processing. MINC data can contain signed and unsigned integer, float and complex data types accompanied by a prepackaged extensible binary format header containing all relevant information regarding the corresponding patient, tumor region and cancer type. Group three of the BITE dataset contains images related to 14 patients all of which were identified to have brain cancer, see Figure.~\ref{fig:anatomical} above. Unfortunately, no pre-operative scan was performed on patient number 13 of the dataset meaning that no manual binary map of the ROI was manually extracted as the patient's pre-operative scan does not exist, hence, patient 13 will be excluded from the future analysis that this paper will propose.

To extract meaningful information from this dataset we primarily used the programming language \texttt{python}\footnote{van Rossum, G., 1990. Python. www.python.org} coupled with neuroimaging libraries such as \texttt{Nibabel}\footnote{Markiewicz, C., 2006. Nibabel. www.nipy.org/nibabel NiPy.} and \texttt{Nilearn}\footnote{2019. Nilearn. https://nilearn.github.io NiPy.} which allowed for easy manipulation of the images. \texttt{Nibabel} was used to load the images in as \texttt{Nibabel} objects which could then be easily converted into \texttt{numpy} arrays to manipulate the images affine matrix. \texttt{Nilearn} a popular \texttt{python} neuroimaging library comes pre-packaged with plots suitable for the visualisation of MR images. Figure.~\ref{fig:anatomical} shows a basic anatomical plot to show the three image perspective planes of a specific MR image. This plot can be altered to visualise different slices of each image plane by manipulating the coordinates of the MR image. By manipulating the coordinates of the image, plots can be designed around the images shape where functions intake a basic slice number as a parameter and increment through the slices of the image displaying each slice as the function increments through the \texttt{numpy} array.

The ground truth accompanying this dataset has several issues that can impact the segmentation performance an algorithm may propose. Most notably, the binary mask of each patient varies in image resolution to the patient's pre-operative scan. The resolution of each binary mask does not match the resolution of each pre or post-operative image. Why does this difference in resolution matter? U-Net intakes an MR two-dimensional image and outputs a segmentation of the ROI supplied to the network. The ROI and the output of the network must be the same resolution of the input image. With the difference in resolution, we are unable to utilize either of these networks until the resolution of the tumor masks is resolved. The difference in image resolution is due to the images being stored in MINC format. This format enables analysts to store cropped regions without losing any coordinate information, making it possible to overlay the full pre-operative MR image with its corresponding small tumor mask without suffering any error in alignment. To overlay a tumor mask with a pre-operative MRI scan we can use one of \texttt{Nilearns} built-in plotting functions to create an ROI plot.

\begin{figure}
    \centering
      \includegraphics[scale=0.6]{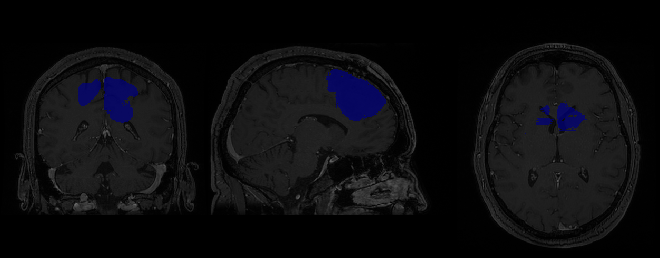}
      \caption{ROI overlay plot, showing patient 2's binary mask overlaid onto their pre-operative scan showing the size, shape and location of their tumor.}
    \label{fig:overlay}
\end{figure}

Figure.~\ref{fig:overlay} represents an ROI plot, where the tumor mask has been overlaid onto the corresponding pre-operative scan. This figure provides us with crucial information regarding the size, shape and location of the patient's tumor. Using \texttt{Nilearn} we can resample the mask to a target dimension. A useful feature which expands our original binary mask to match the image dimensions of each pre-operative scan. The re-sampled mask now meets its corresponding pre-operative mask in terms of height and width, however, this function also reshapes the MINC files slices. As patient 1 has 29 slices compared to 180 the function will add the extra slices to the binary mask so that each scan can be equal. Thus, with this new re-sampled dataset we can begin to explore tumor segmentation using relevant algorithms. Most notably, U-Net works with two-dimensional data. To reformat the MINC data we can extract these 180 slices as two-dimensional images, where each slice represents one image. These slices are then classified as the three perspective planes of the brain, ie., coronal, sagittal, and transversal. Repeating this procedure on 13 patients will create a large dataset for our analysis.


\section{Methodology}\label{sec:methods}

\subsection{Method}\label{sec:method}

U-Net is a fully connected CNN used for efficient semantic segmentation of images. Such U-Net deep neural network fits in various analytical tasks of wide ranging application. This is particularly useful where the input data is the form of images. This architecture has several applications ranging from consumer videos~\cite{dev2019localizing,dev2019identifying}, earth observations~\cite{dev2019multi} and medical imaging~\cite{yin2022u,deng2022elu,ali2022implementation}. The U-Net architecture is based on an autoencoder network where the network will copy its inputs to its outputs~\cite{ronneberger2015u}. An autoencoder network functions by compressing the input image into a latent-space representation which is simply a compressed representation of the images indicating which data points are closest together. The compressed data is later reconstructed to produce an output. An autoencoder network contains two paths, an encoder and a decoder. The encoder compresses the data into a latent-space representation while the decoder is used for the reconstruction of the input data from its latent-space representation. U-Net uses a convolutional autoencoder architecture where the convolutional layers are used to encode and decode the input images.

\begin{figure}
    \centering
      \includegraphics[width=\textwidth]{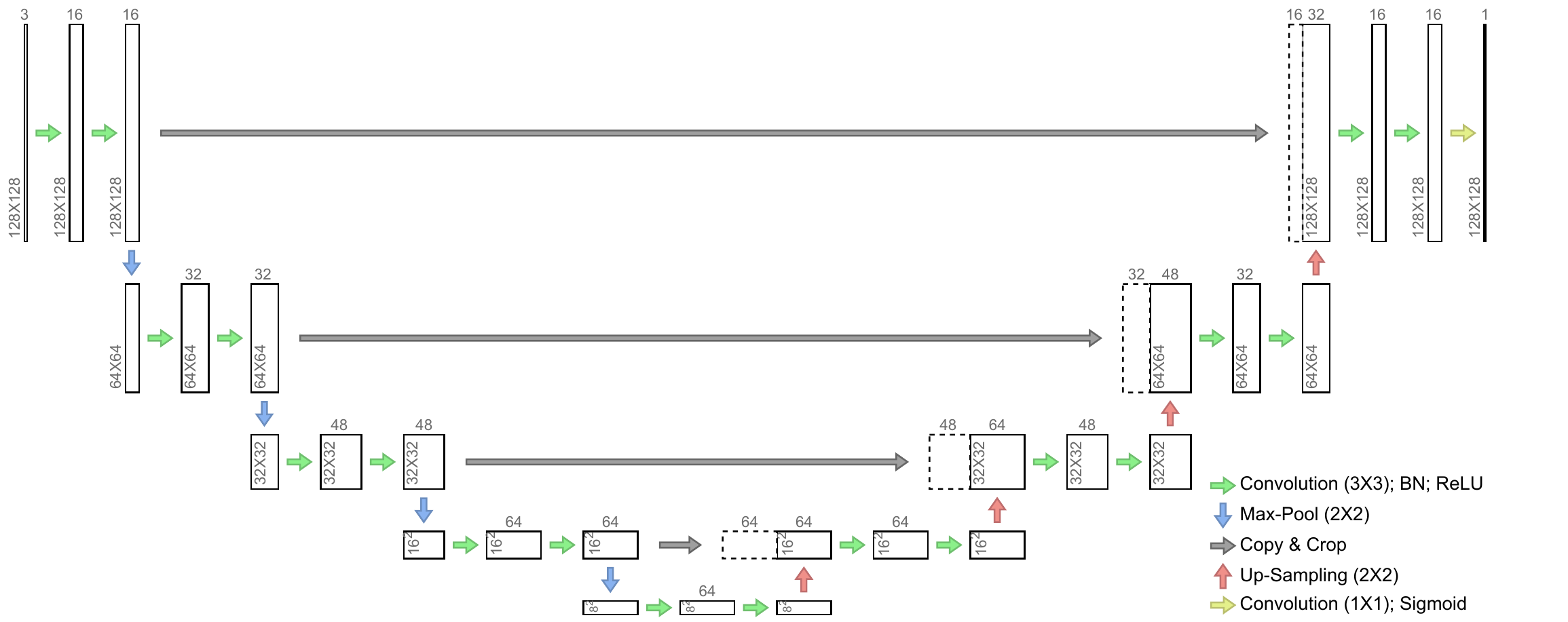}
      \caption{Proposed lightweight implementation of the U-Net architecture.}
    \label{fig:unet}
\end{figure}

Similarly to an autoencoder network, U-Net contains two paths, a contraction path (encoder) and a symmetric expanding path (decoder). The encoder path of U-Net captures the context of the input image, this path is simply a pipeline of convolutional and pooling layers. The decoder path uses transposed convolutions enabling precise localization. 
There is no fully connected feedforward layer (or dense layer) in the U-Net, and it only contains the stacks of convolutional layers and max-pooling layers. Although U-Net was originally designed for $572\times572$ images, it can be easily modified to work with any image dimension~\cite{ronneberger2015u}. Several stacked convolutional layers can enable the network to learn more precise features from the compressed input images, see Figure.~\ref{fig:unet}~\cite{UNET}.

\bigskip

U-Net operates on the assumption that the input image and the 
corresponding binary map are equal in image dimensions. In other words, if the input image is of shape $(512, 512)$, the corresponding binary mask must match this shape.
The input image is compressed to fit into a latent-space representation. The decoder will reconstruct this compressed image back into its original shape of $(512, 512$). U-Net's performance depends solely on the quality of its input image. U-Nets segmentation performance can be evaluated by monitoring its warping, rand and pixel error. Originally, U-Net outperformed a sliding-window convolutional network and produced the best warping error in the EM segmentation challenge. Later iterations of U-Net expanded on U-Nets performance in image segmentation~\cite{ronneberger2015u}.

A very recent addition was U-Net++ which modified U-Net architecture to include a series of nested dense skip pathways, which is used to reduce the gap between feature maps and the pathways (encoder \& decoder) of the network ~\cite{zhou2018unet++}. U-Net++ was also proposed to work with deep supervision. Basically, the loss is now calculated at interim levels alongside the traditional output layers. This has been noted to help with the problem of vanishing gradients during loss backpropagation~\cite{lee2015deeply}.

As U-Net works with two-dimensional data, the acquired dataset was converted from a three-dimensional image plane to a two-dimensional dataset using a `slice extractor' 
The extractor extracts each slice of an MRI scan (MNC file) and saves the slice as a PNG file. Using the extractor we created four new datasets from the original $14$ patients MR images. Three of these datasets will correspond to the perspective planes of the brain (Coronal, Sagittal, Transversal) while the fourth dataset will contain all the available images which adds up to $846$ images in total.


Our proposed implementation of U-Net was achieved using \texttt{Tensorflow}\footnote{Google Brain, $2015$. Tensorflow. www.tensorflow.org/}. Where the network has $4$ convolutional blocks. Each convolutional block of the network contains $2$ convolutional layers with a kernel size of $3\times3$ and zero padding at each layer to control the shrinkage of the object dimension after applying filters. The filter size per convolutional block varies after each layer where the filter size increments in steps of 16. Each layer of a convolutional block is activated by a Rectified Linear Unit (ReLU) while in between these layers a batch normalization step is then applied. At the encoder layer of the network, we apply a $2\times2$ max pooling layer after a function call to add a convolutional block to further reduce the spatial dimensions of the input image. While max pooling is also applied at the decoder layer its application here is too up-sample the feature map using the memorized max-pooling indices~\cite{ronneberger2015u}.


\subsection{Training \& Optimization}

During the training process, we decided to stick with a simple cost function for this network. We chose binary cross-entropy loss (say, $L_{BCE}$) for our cost function as the network is simply being trained to segment one particular region, the cancerous section of an input MR image. This loss function expects a sigmoid outcome ($\hat{y}_{i}$) as it is a binary predictor with target values ($y_{i}$) as $1$ or $0$. Given the output size of $N$, $L_{BCE}$ is defined as per equation~\ref{eq:lossBCE}. Prior consideration was taken to decide on fixed image size for our inputs. As the image size varies across the three perspective planes we decided on an image size of $128 \times 128$ as the images extracted from both the coronal and sagittal planes are substantially smaller than those extracted from the transversal plane. With this image size in mind, the networks architecture was designed to be lightweight and swift so a prediction image could be reproduced. This image size was chosen based on the variation of dimensions of the original patient's image data.

\begin{equation}
    L_{BCE} = \frac{1}{N}\sum_{i=1}^{N}\left[y_{i}\cdot\left(\log \hat{y}_{i}\right) + \left(1-y_{i}\right)\cdot\left(log\left(1-\hat{y}_{i}\right)\right)\right]
    \label{eq:lossBCE}
\end{equation}

Before the training of each network could start, metric callbacks were introduced to control the performance of the network. \texttt{Early Stopping} and \texttt{Model Checkpointing} were used to ensure the performance of the network did not degrade if extreme values were introduced into the networks epoch range. 
To monitor the performance of the network, precision, recall and 
intersection-over-union (IoU) was logged using \texttt{CSV Logger}. This log helped us to analyze the stepwise results that were produced by the network during the training process. The model was trained with Adam optimizer with a learning rate of $0.0001$ which was chosen after rigorous experimentation. 

The epoch range used for training varied based on the results recorded from prior experiments. We decided that running small experiments with only $10$ epochs would allow us to quickly grasp how our network performs per perspective plane. After a small number of iterations were performed we gradually increased this parameter range to $50$ epochs to monitor the networks convergence rate. One particular parameter, we also experimented with, was the networks filter size. By increasing the networks number of filters we are essentially increasing the number of trainable parameters in the model. This particular experiment was undertaken to observe the performance increase when the total trainable parameters of the network are increased via the convolution blocks filter range. Several iterations were undertaken using a variety of filter values for the networks convolutional block. The optimization was done by reflecting on the stepwise IoU values. During the experiments, filter values were incremented by $16$ per convolutional block after each iteration.


\section{Results}\label{sec:results}

\subsection{Subjective Evaluation}\label{sec:subjectiveEvaluation}

\begin{figure}
    \centering
      \includegraphics[scale=0.6]{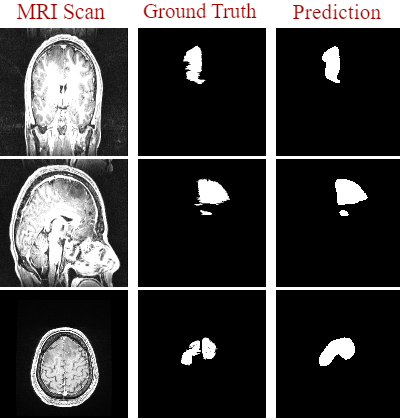}
      \caption{Segmentation results recorded using U-Net with 50 epochs.}
    \label{fig:results_1}
\end{figure}

Training the network began in small iterations where we monitored the segmentation performance based on iterations alone. Early results using $10$ epochs produced poor results across two of the image planes notably the sagittal and coronal planes. Poor segmentation results were expected based on the small epoch range used during training, however, adequate results were recorded on the transversal plane. This is likely due to the size of each dataset as the transversal dataset had the most images as the patient's tumor is most prominent from this particular perspective. To further improve these results we increased the networks epoch range to $50$ and monitored the results to see if the segmentation performance had improved. Using $50$ epochs significantly improved the networks segmentation performance across all three perspective planes. Figure.~\ref{fig:results_1} shows the results across all three perspectives using only $50$ epochs and the proposed network architecture from Section~\ref{sec:method}. Here, model 1 (or first model) is the standard U-Net architecture, whereas the model 2 (or the final model) is the proposed U-Net with optimized filter values.

To further improve the networks segmentation performance we began undertaking small experiments using the entire dataset of images. This dataset contains all images extracted from all three perspective planes. The purpose of this study is to experiment on U-Nets ability to extract features from images from different perspectives. Running the network using this new dataset for 50 epochs only produced very promising results. Increasing the number of epochs to an extreme size does increase the networks overall segmentation accuracy but only by a very slight amount.

\begin{figure}
    \centering
      \includegraphics[scale=0.3]{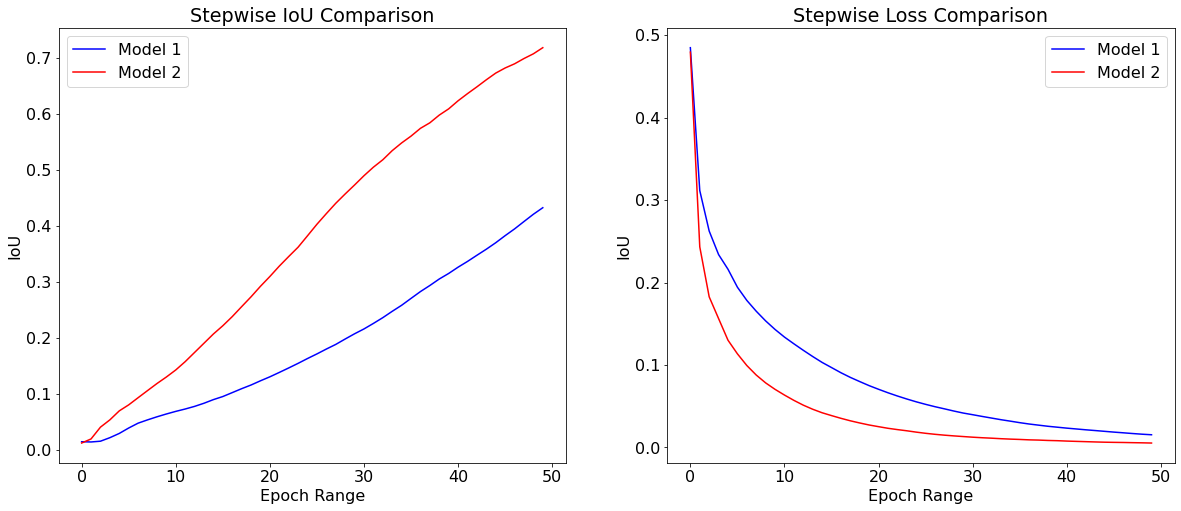}
      \caption{Direct comparison of the stepwise IoU and stepwise loss recorded by the first and final model during the training phase.}
    \label{fig:results_iouLoss}
\end{figure}

Overall, we observed that with a very small number of trainable parameters and a very small epoch size, our network can provide accurate segmentation of the cancerous region of a T1-weighted MR image. Using a small filter range while training on the entire dataset, the network produced a final stepwise IoU of $43\%$. To improve the networks segmentation performance we performed several experiments on the convolutional block's filter range. As previously mentioned we incremented via steps of 16 and monitored the stepwise IoU after each epoch until each training iteration was complete. The results from this experiment which can be observed in Figure.~\ref{fig:results_iouLoss} drastically improved the networks IoU and also allowed for a faster convergence rate of our models loss.

\begin{figure}
    \centering
      \includegraphics[scale=0.8]{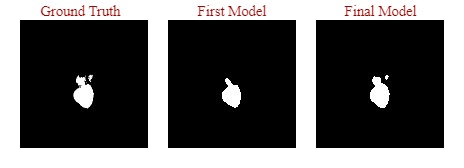}
      \caption{Comparison of results from U-Net by increasing the number of trainable parameters the network contains.}
    \label{fig:results_2}
\end{figure}

By increasing the filter values three times and by training the network with $50$ epochs using the entire dataset we recorded a direct increase in the networks segmentation performance and setpwise IoU. The networks stepwise IoU had drastically increased to $71\%$ which is significantly higher then the previous results recorded using the standard network. By observing Figure.~\ref{fig:results_2} we can see the effects of an increased filter range on our proposed network. We observed that an increased filter value increased the networks segmentation performance by providing a more accurate structure to the network's segmentation results. Thus, using a larger filter range allows our implementation of U-Net to extract more meaningful features from the input images on either of the four proposed datasets.

\subsection{Objective Evaluation}

With the results from our proposed network, we performed basic predictions on the withdrawn test set ($10\%$ of the total number of images) of each of the four datasets. Each prediction was resized and extracted to a separate location indicated by its perspective plane or image contents. With these new predictions from our proposed network, we created implementations of four different popular image processing evaluation metrics to evaluate the overall quality of these predictions in comparison to the supplied ground truth images. These four evaluation metrics~\cite{batra2022dmcnet}, namely, pixel accuracy, mean accuracy, mean IoU and frequency weighted IoU (FWIoU) can be calculated using the following formulas:
\bigskip

\begin{equation}
    \mbox{Pixel Accuracy} = \sum_i= \frac{n_{ii}}{\sum_i t_i}
\end{equation}
\bigskip

\begin{equation}
    \mbox{Mean Accuracy} = (1/n_{cl})\sum_i \frac{n_{ii}}{t_i}
\end{equation}
\bigskip

\begin{equation}
    \mbox{Mean IoU} = (1/n_{cl})\sum_i \frac{n_{ii}}{t_i + \sum_i n_{ji} - n_{ii}}
\end{equation}
\bigskip

\begin{equation}
    \mbox{FWIoU} = (\sum_k t_k)^{-1} \sum_i \frac{t_i n_{ii}}{t_i + \sum_j n_{ji} - n_{ii}}
\end{equation}
\bigskip

where the value $n_{ij}$ represents the number of pixels within class $i$ which was classified to class $j$. The value $n_{cl}$ symbolizes the total number of classes, which in this case is only 1 as this is not a multi-class segmentation problem. Finally, the value of $t_i$ represents the total number of pixels in class $i$~\cite{liu2019recent}.

Table.~\ref{table1} below shows the results obtained using our first implementation of U-Net trained on 50 epochs and evaluated on each withheld test set corresponding to each perspective plane. The results from Table.~\ref{table1} show that our first implementation of U-Net which had a very small number of trainable parameters yielded an excellent average mean IoU score of 70\% - 80\% across all three of perspective planes, additionally clarified in Section~\ref{sec:subjectiveEvaluation} we used an additional dataset containing the entire collection of brain scans regardless of their perspective. The results from this dataset labelled by the heading 'Full' show that our network accumulated a mean IoU of 84\%. A high mean IoU for this dataset was expected, however, the slight drop in accuracy related to the perspective planes is likely due to the size of the dataset which the network was trained on. Data augmentation was not utilized based on these results alone. This is because the network could already produce accurate segmentation results without the interference of a data augmentation pipeline. This is in contrast with the original implementation of U-Net which was trained on a very small number of images but required a high degree of data augmentation.

\begin{table}
\centering
\caption{Results obtained during our first iteration of training using 50 epochs with a small filter range per convolutional block.}\label{table1}
\resizebox{\textwidth}{!}{
    \begin{tabular}{|l|c|c|c|c|}
        \hline 
        \textbf{Dataset} &  \textbf{Pixel Acc (\%)} & \textbf{Mean Acc (\%)} & \textbf{Mean IoU (\%)} & \textbf{FWIoU (\%)} \\
        \hline \hline
        Coronal & 99 & 82 & 81 & 99 \\
        Sagittal & 99 & 74 & 72 & 99 \\
        Transversal & 99 & 82 & 77 & 99\\
        Full & 99 & 87 & 84 & 99\\
        \hline
    \end{tabular}}
\end{table}

However, an indirect comparison to the results can be observed from Table.~\ref{table2}. We can observe that by increasing the total number of trainable parameters of our network trained using the entirety of the images, the mean IoU score directly increases by $5\%$,. Improvements were also observed across all three of the perspective planes where the average mean IoU score had increased by a range of $4\% - 5\%$. Thus, we can conclude that the increase of the number of trainable parameters of the network has a positive impact on the segmentation performance of the network.

\begin{table}
\centering
\caption{Results obtained during our second iteration of training using 50 epochs with a significantly larger filter range per convolutional block.}\label{table2}
    \resizebox{\textwidth}{!}{\begin{tabular}{|l|c|c|c|c|}
        \hline
        \textbf{Dataset} &  \textbf{Pixel Acc (\%)} & \textbf{Mean Acc (\%)} & \textbf{Mean IoU (\%)} & \textbf{FWIoU (\%)} \\
        \hline \hline
        Coronal & 99 & 88 & 84 & 99 \\
        Sagittal & 99 & 76 & 75 & 99 \\
        Transversal & 99 & 85 & 84 & 99\\
        Full & 99 & 91 & 89 & 99\\
        \hline
    \end{tabular}}
\end{table}

\subsection{Comparison of proposed method with other methods}

\begin{figure}
    \centering
      \includegraphics[scale=0.8]{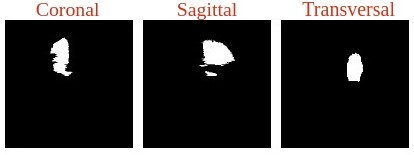}
      \caption{The ground truth images used to evaluate our chosen benchmarking algorithms.}
    \label{fig:ground_truth}
\end{figure}

\begin{figure}
    \centering
      \includegraphics[scale=0.5]{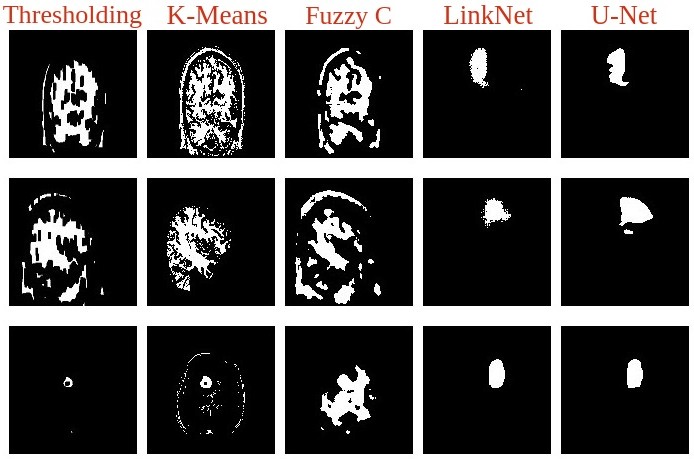}
      \caption{Comparison of the results obtained during the benchmarking process to those predicted via U-Net.}
    \label{fig:benchmark_1}
\end{figure}

Several algorithms were chosen to benchmark to further evaluate the segmentation performance obtained using U-Net. These algorithms specified in Section~\ref{sec:segTech} of the paper were chosen based on their success on biomedical image segmentation. Each algorithm was used to compare our segmentation results obtained using our proposed model. Each algorithm was tested using the extracted test sets of each of the four available datasets. Figure~\ref{fig:ground_truth} shows the three ground truth images which correspond directly to the predictions generated in Figure~\ref{fig:benchmark_1}.

Comparing the results obtained using U-Net to those obtained to benchmark, we observed that our proposed model outperformed each of these benchmarking algorithms. While the deep learning methods perform segmentation task automatically, the other three benchmarking algorithms (Thresholding, K-Means, Fuzzy C-Means) needed more `manual' interference. They perform better when the tumor is clearly distinguishable at the pixel level in terms of intensity and isolation. This also means that to achieve maximum performance when the parameters are fine-tuned separately for each image, which essentially renders the whole process useless. 
The three-manual segmentation algorithms were implemented for automatic segmentation of brain tumors from MRI images. However, these algorithms are not fully automatic and may require fine-tuning for good results. Thus, a lightweight deep-learning architecture which can provide real-time automatic segmentation of brain tumors is needed.

More promising results came from LinkNet which can be observed in Table~\ref{table3} below. The implementation of LinkNet was consistent with the original architecture used when it was first proposed. However, small changes were made to the network to suit our segmentation goal. These small changes were simply related to the particular activation function used at the final layer of the network. Also, the common loss function for multi-class segmentation problems, which was originally used (\textit{i.e.} categorical cross-entropy), was changed. Visual predictions generated using LinkNet were adequate. However, the performance fluctuates as the perspective of the image changes, for instance in Figure.~\ref{fig:benchmark_1} on the coronal and sagittal plane we observed mixed results as the network successfully captures the location of the tumor but cannot directly generate its size or structure. Whereas on the transversal plane, comparing the generated prediction with its corresponding ground-truth from Figure.~\ref{fig:ground_truth}, we observe an increase in segmentation performance as the prediction is very accurate. This performance increase is likely due to the size of the dataset as the transversal plane contains more images than both the coronal and sagittal planes combined.

\begin{table}
\centering
\caption{Comparison of the results obtained during the benchmarking process.}\label{table3}
   \resizebox{\textwidth}{!}{ \begin{tabular}{|l|c|c|c|c|}
         \hline
        \textbf{Method} &  \textbf{Pixel Acc (\%)} & \textbf{Mean Acc (\%)} & \textbf{Mean IoU (\%)} & \textbf{FWIoU (\%)} \\ 
        \hline \hline
      
        \multicolumn{5}{|c|}{\textbf{Coronal}} \\
        \hline
        Thresholding & 91 & 70 & 47 & 90 \\
        K-Means & 79 & 72 & 41 & 78 \\
        Fuzzy C & 84 & 73 & 44 & 83\\
        LinkNet & 99 & 78 & 76 & 99\\
        U-Net & \textbf{99} & \textbf{88} & \textbf{84} & \textbf{99}\\
        \hline
        \multicolumn{5}{|c|}{\textbf{Sagittal}} \\
        \hline
        Thresholding & 93 & 58 & 48 & 92 \\
        K-Means & 82 & 60 & 42 & 81 \\
        Fuzzy C & 86 & 56 & 44 & 85\\
        LinkNet & 99 & 60 & 59 & 98\\
        U-Net & \textbf{99} & \textbf{76} & \textbf{75} & \textbf{99}\\
        \hline
        \multicolumn{5}{|c|}{\textbf{Transversal}} \\
        \hline
        Thresholding & 97 & 59 & 51 & 96 \\
        K-Means & 91 & 67 & 48 & 90 \\
        Fuzzy C & 91 & 70 & 48 & 91\\
        LinkNet & 99 & 83 & 81 & 99\\
        U-Net & \textbf{99} & \textbf{85} & \textbf{84} & \textbf{99}\\
        \hline
        \multicolumn{5}{|c|}{\textbf{Full}} \\
        \hline
        Thresholding & 95 & 61 & 49 & 93 \\
        K-Means & 86 & 65 & 45 & 85 \\
        Fuzzy C & 88 & 65 & 45 & 87\\
        LinkNet & 99 & 87 & 84 & 99\\
        U-Net & \textbf{99} & \textbf{91} & \textbf{89} & \textbf{99}\\
        \hline
    \end{tabular}}
\end{table}

\begin{figure}
    \centering
      \includegraphics[scale=0.8]{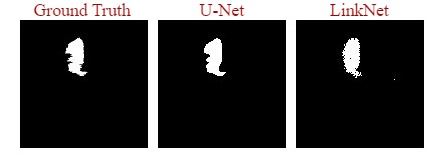}
      \caption{Comparison of the performance difference obtained via LinkNet and our proposed model trained on 50 epochs.}
    \label{fig:benchmark_2}
\end{figure}

LinkNet was trained for $50$ epochs to match the training phase of the proposed implementation of U-Net. The proposed method yielded a mean IoU which was $4\%$ higher than that recorded with LinkNet when trained on the entire dataset. Improvements were also recorded on all the three perspective planes. We can visualize this difference using our evaluation metrics by comparing the predictions generated by the two networks. For instance, in Figure.~\ref{fig:benchmark_2} we observe the segmentation of U-Net when compared to the prediction generated by LinkNet. The predictions are both very accurate, however, there is a $10\%$ mean accuracy difference between these two models. This difference in accuracy shows how LinkNet produces a correct prediction but cannot define the structure of its prediction on the small epoch range which it was trained on. Our proposed model outperformed each of the four widely used algorithms on all four evaluation metrics. This means that the model we have proposed is a lightweight network which is also very accurate at segmenting brain anomalies from MRI two-dimensional scans.

\subsection{Discussion}

The purpose of these experiments was to explore how a lightweight network could perform real-time automatic image segmentation on MRI brain scans. Image segmentation on medical images can provide accurate properties such as the size, shape and location of an apparent mass discovered by medical physicians. As previously discussed in earlier sections of this paper, our implementation of U-Net is a lightweight version trained specifically to segment brain tumors from MRI brains scans. Performing several training iterations allowed us to gradually fine-tune the network so we could monitor and improve U-Net's performance on each perspective plane and the entire dataset as a whole. By undertaking such a cautious approach to training it allowed us to quickly observe the convergence and overall performance our network produced, thus there was an unessential need for data augmentation.

The use of systematic benchmarking allowed us to compare our model's performance to four other widely used methods in the domain of image processing. These methods although not entirely suited to medical image segmentation were previously used in applications where either a single or multi-class segmentation occurred. By including both manual and automatic segmentation techniques we could directly compare the performance obtained using our proposed method to that obtained by one of the four widely used benchmarking methods. These experiments revealed that our proposed method outperformed each of the four methods used for benchmarking on each of the perspective planes and the entirety of the dataset, indicating that the network we propose is incredibly accurate at segmenting brain tumors from two-dimensional MRI brain scans.


\section{Conclusion}\label{sec:conc}

Due to the purpose of our study, we have seen how U-Net an existing deep learning architecture for biomedical image segmentation can be altered and fine-tuned for brain tumor segmentation. By using the BITE's dataset and converting these three-dimensional MRI brain scans to two-dimensional images we have been able to use this data to evaluate the performance of a very lightweight implementation of U-Net which can accurately segment anomalies from two-dimensional images. The network outperforms any of the standard benchmarking algorithms used to evaluate the performance of the network, the network yields an average mean IoU of 84\% when trained on the entire dataset, interestingly the mean IoU does not stagnate when the network is trained only on one the three perspective planes, a study which was undertaken to observe U-Nets approach to segmenting anomalies on small datasets containing less than one hundred images.

Our implementation of U-Net is lightweight and can perform accurate segmentation's, without the need for aggressive data augmentation. This proposed network could be used in a medical setting for trained physicians to have a second evaluator to a patients MR image. Research on the particular topic of brain tumor segmentation has advanced rapidly with the application of deep learning, however, more studies are needed to further improve the performance of a proposed network as the ratio between the predicted images False Negatives and False Positives is crucial in biomedical image analysis. We intend to benchmark our proposed lightweight U-Net with the original U-Net structure and other statistical~\cite{dev2018high} and deep learning networks~\cite{minaee2021image}. Future work could improve on this performance by investigating the use of data augmentation to artificially increase the size of the dataset using an augmentor pipeline. Nevertheless, this study has shown how deep learning and computer vision can be applied in a medical domain to accurately segment brain tumors from two-dimensional MR brain images using a lightweight variant of a well known architecture.


\section*{Acknowledgments}
This research was conducted with the financial support of Science Foundation Ireland under Grant Agreement No. 13/RC/2106\_P2 at the ADAPT SFI Research Centre at University College Dublin. ADAPT, the SFI Research Centre for AI-Driven Digital Content Technology, is funded by Science Foundation Ireland through the SFI Research Centres Programme.


\begin{thebibliography}{10}
\expandafter\ifx\csname url\endcsname\relax
  \def\url#1{\texttt{#1}}\fi
\expandafter\ifx\csname urlprefix\endcsname\relax\def\urlprefix{URL }\fi
\expandafter\ifx\csname href\endcsname\relax
  \def\href#1#2{#2} \def\path#1{#1}\fi

\bibitem{patel2020benign}
A.~Patel, Benign vs malignant tumors, JAMA oncology 6~(9) (2020) 1488--1488.

\bibitem{icsin2016review}
A.~I{\c{s}}{\i}n, C.~Direko{\u{g}}lu, M.~{\c{S}}ah, Review of mri-based brain
  tumor image segmentation using deep learning methods, Procedia Computer
  Science 102 (2016) 317--324.

\bibitem{md2022automatic}
A.~Md.~Sattar, M.~Kr.~Ranjan, Automatic cancer detection using probabilistic
  convergence theory, in: Computational Intelligence in Oncology: Applications
  in Diagnosis, Prognosis and Therapeutics of Cancers, Springer, 2022, pp.
  111--122.

\bibitem{pathan2022analyzing}
M.~S. Pathan, A.~Nag, M.~M. Pathan, S.~Dev, Analyzing the impact of feature
  selection on the accuracy of heart disease prediction, Healthcare Analytics 2
  (2022) 100060.

\bibitem{nwosu2019predicting}
C.~S. Nwosu, S.~Dev, P.~Bhardwaj, B.~Veeravalli, D.~John, Predicting stroke
  from electronic health records, in: 2019 41st Annual International Conference
  of the IEEE Engineering in Medicine and Biology Society (EMBC), IEEE, 2019,
  pp. 5704--5707.

\bibitem{dev2022predictive}
S.~Dev, H.~Wang, C.~S. Nwosu, N.~Jain, B.~Veeravalli, D.~John, A predictive
  analytics approach for stroke prediction using machine learning and neural
  networks, Healthcare Analytics 2 (2022) 100032.

\bibitem{pathan2020identifying}
M.~S. Pathan, Z.~Jianbiao, D.~John, A.~Nag, S.~Dev, Identifying stroke
  indicators using rough sets, IEEE Access 8 (2020) 210318--210327.

\bibitem{sivapalan2022annet}
G.~Sivapalan, K.~K. Nundy, S.~Dev, B.~Cardiff, D.~John, {ANNet}: a lightweight
  neural network for {ECG} anomaly detection in {IoT} edge sensors, IEEE
  Transactions on Biomedical Circuits and Systems 16~(1) (2022) 24--35.

\bibitem{medical_imaging}
M.~I. Modalities, \href{https://bit.ly/37Ci6R7}{The three perspective planes of
  the brain.}, 2013.
\newline\urlprefix\url{https://bit.ly/37Ci6R7}

\bibitem{hu2022mutual}
J.~Hu, X.~Gu, X.~Gu, Mutual ensemble learning for brain tumor segmentation,
  Neurocomputing 504 (2022) 68--81.

\bibitem{liu2021review}
X.~Liu, L.~Song, S.~Liu, Y.~Zhang, A review of deep-learning-based medical
  image segmentation methods, Sustainability 13~(3) (2021) 1224.

\bibitem{ronneberger2015u}
O.~Ronneberger, P.~Fischer, T.~Brox, U-net: Convolutional networks for
  biomedical image segmentation, in: International Conference on Medical image
  computing and computer-assisted intervention, Springer, 2015, pp. 234--241.

\bibitem{li2018lightweight}
Y.~Li, J.~Liu, L.~Wang, Lightweight network research based on deep learning: A
  review, in: 2018 37th Chinese control conference (CCC), IEEE, 2018, pp.
  9021--9026.

\bibitem{ali2022implementation}
O.~Ali, H.~Ali, S.~A.~A. Shah, A.~Shahzad, Implementation of a modified u-net
  for medical image segmentation on edge devices, IEEE Transactions on Circuits
  and Systems II: Express Briefs (2022).

\bibitem{dev2019cloudsegnet}
S.~Dev, A.~Nautiyal, Y.~H. Lee, S.~Winkler, Cloudsegnet: A deep network for
  nychthemeron cloud image segmentation, IEEE Geoscience and Remote Sensing
  Letters 16~(12) (2019) 1814--1818.

\bibitem{dev2016color}
S.~Dev, Y.~H. Lee, S.~Winkler, Color-based segmentation of sky/cloud images
  from ground-based cameras, IEEE Journal of Selected Topics in Applied Earth
  Observations and Remote Sensing 10~(1) (2016) 231--242.

\bibitem{jain2021using}
M.~Jain, C.~Meegan, S.~Dev, Using {GANs} to augment data for cloud image
  segmentation task, in: {2021 IEEE International Geoscience and Remote Sensing
  Symposium (IGARSS)}, IEEE, 2021, pp. 3452--3455.

\bibitem{minaee2021image}
S.~Minaee, Y.~Y. Boykov, F.~Porikli, A.~J. Plaza, N.~Kehtarnavaz,
  D.~Terzopoulos, Image segmentation using deep learning: {A} survey, IEEE
  transactions on pattern analysis and machine intelligence (2021).

\bibitem{bien2018deep}
N.~Bien, P.~Rajpurkar, R.~L. Ball, J.~Irvin, A.~Park, E.~Jones, M.~Bereket,
  B.~N. Patel, K.~W. Yeom, K.~Shpanskaya, et~al., Deep-learning-assisted
  diagnosis for knee magnetic resonance imaging: development and retrospective
  validation of mrnet, PLoS medicine 15~(11) (2018) e1002699.

\bibitem{zhou2018unet++}
Z.~Zhou, M.~M.~R. Siddiquee, N.~Tajbakhsh, J.~Liang, Unet++: A nested u-net
  architecture for medical image segmentation, in: Deep Learning in Medical
  Image Analysis and Multimodal Learning for Clinical Decision Support,
  Springer, 2018, pp. 3--11.

\bibitem{pham2000current}
D.~L. Pham, C.~Xu, J.~L. Prince, Current methods in medical image segmentation,
  Annual review of biomedical engineering 2~(1) (2000) 315--337.

\bibitem{Otsu1979}
N.~Otsu, A threshold selection method from gray-level histograms, IEEE
  transactions on systems, man, and cybernetics 9~(1) (1979) 62--66.

\bibitem{mustaqeem2012efficient}
A.~Mustaqeem, A.~Javed, T.~Fatima, An efficient brain tumor detection algorithm
  using watershed \& thresholding based segmentation, International Journal of
  Image, Graphics and Signal Processing 4~(10) (2012) 34.

\bibitem{ho2021multivariate}
Z.~Y. Ho, M.~Jain, S.~Dev, Multivariate Convolutional LSTMs for Relative
  Humidity Forecasting, 2021, pp. 2317--2323.

\bibitem{dev2019identifying}
S.~Dev, H.~Javidnia, M.~Hossari, M.~Nicholson, K.~McCabe, A.~Nautiyal,
  C.~Conran, J.~Tang, W.~Xu, F.~Piti{\'e}, Identifying candidate spaces for
  advert implantation, 2019, pp. 503--507.

\bibitem{sharma2019improved}
S.~Sharma, M.~Rattan, An improved segmentation and classifier approach based on
  hmm for brain cancer detection, The Open Biomedical Engineering Journal
  13~(1) (2019).

\bibitem{mercier2012online}
L.~Mercier, R.~F. Del~Maestro, K.~Petrecca, D.~Araujo, C.~Haegelen, D.~L.
  Collins, Online database of clinical mr and ultrasound images of brain
  tumors, Medical Physics 39~(6-Part1) (2012) 3253--3261.

\bibitem{jain2022internal}
M.~Jain, M.~Jain, T.~AlSkaif, S.~Dev, Which internal validation indices to use
  while clustering electric load demand profiles?, Sustainable Energy, Grids
  and Networks 32 (2022) 100849.

\bibitem{jain2020clustering}
M.~Jain, T.~AlSkaif, S.~Dev, A clustering framework for residential electric
  demand profiles, 2020, pp. 1--6.

\bibitem{jain2021validation}
M.~Jain, T.~AlSkaif, S.~Dev, Validating clustering frameworks for electric load
  demand profiles, IEEE Transactions on Industrial Informatics 17~(12) (2021)
  8057--8065.
\newblock \href {https://doi.org/10.1109/TII.2021.3061470}
  {\path{doi:10.1109/TII.2021.3061470}}.

\bibitem{dev2014systematic}
S.~Dev, Y.~H. Lee, S.~Winkler, Systematic study of color spaces and components
  for the segmentation of sky/cloud images, 2014, pp. 5102--5106.

\bibitem{kabade2013segmentation}
R.~S. Kabade, M.~Gaikwad, Segmentation of brain tumour and its area calculation
  in brain mr images using k-mean clustering and fuzzy c-mean algorithm,
  International Journal of Computer Science \& Engineering Technology 4~(05)
  (2013) 524--531.

\bibitem{abdel2015brain}
E.~Abdel-Maksoud, M.~Elmogy, R.~Al-Awadi, Brain tumor segmentation based on a
  hybrid clustering technique, Egyptian Informatics Journal 16~(1) (2015)
  71--81.

\bibitem{wang2022amdcnet}
H.~Wang, Y.~Li, S.~Xi, S.~Wang, M.~S. Pathan, S.~Dev, {AMDCNet}: An attentional
  multi-directional convolutional network for stereo matching, Displays (2022)
  102243.

\bibitem{chaurasia2017linknet}
A.~Chaurasia, E.~Culurciello, Linknet: Exploiting encoder representations for
  efficient semantic segmentation, in: 2017 IEEE Visual Communications and
  Image Processing (VCIP), IEEE, 2017, pp. 1--4.

\bibitem{BrostowFC:PRL2008}
G.~J. Brostow, J.~Fauqueur, R.~Cipolla, Semantic object classes in video: A
  high-definition ground truth database, Pattern Recognition Letters xx~(x)
  (2008) xx--xx.

\bibitem{clark2013cancer}
K.~Clark, B.~Vendt, K.~Smith, J.~Freymann, J.~Kirby, P.~Koppel, S.~Moore,
  S.~Phillips, D.~Maffitt, M.~Pringle, et~al., The cancer imaging archive
  (tcia): maintaining and operating a public information repository, Journal of
  digital imaging 26~(6) (2013) 1045--1057.

\bibitem{dev2019localizing}
S.~Dev, M.~Hossari, M.~Nicholson, K.~McCabe, A.~Nautiyal, C.~Conran, J.~Tang,
  W.~Xu, F.~Piti{\'e}, Localizing adverts in outdoor scenes, in: 2019 IEEE
  International Conference on Multimedia \& Expo Workshops (ICMEW), IEEE, 2019,
  pp. 591--594.

\bibitem{dev2019multi}
S.~Dev, S.~Manandhar, Y.~H. Lee, S.~Winkler, Multi-label cloud segmentation
  using a deep network, in: 2019 USNC-URSI Radio Science Meeting (Joint with
  AP-S Symposium), IEEE, 2019, pp. 113--114.

\bibitem{yin2022u}
X.-X. Yin, L.~Sun, Y.~Fu, R.~Lu, Y.~Zhang, U-net-based medical image
  segmentation, Journal of Healthcare Engineering 2022 (2022).

\bibitem{deng2022elu}
Y.~Deng, Y.~Hou, J.~Yan, D.~Zeng, Elu-net: An efficient and lightweight u-net
  for medical image segmentation, IEEE Access 10 (2022) 35932--35941.

\bibitem{UNET}
F.~P. Ronneberger~O., B.~T., \href{https://arxiv.org/abs/1505.04597}{U-Net
  Architecture}, 2015.
\newline\urlprefix\url{https://arxiv.org/abs/1505.04597}

\bibitem{lee2015deeply}
C.-Y. Lee, S.~Xie, P.~Gallagher, Z.~Zhang, Z.~Tu, Deeply-supervised nets, in:
  Artificial intelligence and statistics, PMLR, 2015, pp. 562--570.

\bibitem{batra2022dmcnet}
S.~Batra, H.~Wang, A.~Nag, P.~Brodeur, M.~Checkley, A.~Klinkert, S.~Dev,
  {DMCNet}: Diversified model combination network for understanding engagement
  from video screengrabs, Systems and Soft Computing 4 (2022) 200039.

\bibitem{liu2019recent}
X.~Liu, Z.~Deng, Y.~Yang, Recent progress in semantic image segmentation,
  Artificial Intelligence Review 52~(2) (2019) 1089--1106.

\bibitem{dev2018high}
S.~Dev, F.~M. Savoy, Y.~H. Lee, S.~Winkler, High-dynamic-range imaging for
  cloud segmentation, Atmospheric Measurement Techniques 11~(4) (2018)
  2041--2049.

\end{thebibliography}
\end{document}